\newcommand*{\ATLASLATEXPATH}{latex/}
\pdfoutput=1
\documentclass[cernpreprint,texlive=2011,txfonts,UKenglish]{\ATLASLATEXPATH atlasdoc}
\usepackage{\ATLASLATEXPATH atlaspackage}
\usepackage{\ATLASLATEXPATH atlasbiblatex}

\usepackage{\ATLASLATEXPATH atlascontribute}

\usepackage{\ATLASLATEXPATH atlasphysics}
\usepackage{bm}
\graphicspath{{logos/}{figures/}}

\usepackage{PhotonJetResonances2015_1fb-defs}

\usepackage{listings}
\AtlasTitle{Search for new phenomena with photon+jet events in proton–proton collisions at $\sqrt{s} = 13$ TeV with the ATLAS detector}

\author{The ATLAS Collaboration}

 \usepackage{authblk}
 \renewcommand\Authands{, } % avoid ``. and'' for last author
 \renewcommand\Affilfont{\itshape\small} % affiliation formatting
\date{\today}

\AtlasRefCode{EXOT-2015-12}

\AtlasNote{ATL-COM-PHYS-2015-628}
\PreprintIdNumber{CERN-PH-EP-2015-320}

\AtlasJournal{JHEP}
\AtlasAbstract{%
A search is performed for the production of high-mass resonances decaying
into a photon and a jet in \datalumi of proton--proton
collisions at a centre-of-mass energy of $\sqrt{s} = 13$~TeV
collected by the ATLAS detector at the Large Hadron Collider.
Selected events have an isolated photon and a jet, each
with transverse momentum above 150 GeV.
No significant deviation of the $\gamma+$jet invariant mass
distribution from the background-only hypothesis is found.
Limits are set at 95\% confidence level on the cross sections of
generic Gaussian-shaped signals and of a few benchmark phenomena
beyond the Standard Model: excited quarks with vector-like
couplings to the Standard Model particles, and
non-thermal quantum black holes in two models of extra spatial
dimensions.
The minimum excluded visible cross sections for Gaussian-shaped resonances
with width-to-mass ratios of 2\% decrease from about 6~fb
for a mass of 1.5~TeV to about 0.8~fb for a mass of 5 TeV.
The minimum excluded visible cross sections for Gaussian-shaped resonances
with width-to-mass ratios of 15\% decrease from about 50~fb
for a mass of 1.5~TeV to about 1.0~fb for a mass of 5 TeV.
Excited quarks are excluded below masses of 4.4~TeV, and non-thermal
quantum black holes are excluded below masses of 3.8 (6.2)~TeV for
Randall--Sundrum (Arkani-Hamed--Dimopoulous--Dvali) models with
one (six) extra dimensions.
}

\AtlasCoverSupportingNote{Search for new physics in photon+jet final states at 13 TeV}{http://cds.cern.ch/record/2030268}
\AtlasCoverCommentsDeadline{10th December 2015}

\AtlasCoverAnalysisTeam{J. Beringer, L. Carminati, M.-H. Genest, D.M. Gingrich, G. Marchiori, D. Nole, Y. Okumura, R. Turra, M. Villaplana}

\AtlasCoverEdBoardMember{Krisztian~Peters~(chair)}
\AtlasCoverEdBoardMember{Juan~Terron}
\AtlasCoverEdBoardMember{Fernando~Monticelli}

\AtlasCoverEgroupEditors{atlas-EXOT-2015-12-editors@cern.ch}

\AtlasCoverEgroupEdBoard{atlas-EXOT-2015-12-editorial-board@cern.ch}

\hypersetup{pdftitle={ATLAS draft},pdfauthor={The ATLAS Collaboration}}

\begin{document}

\maketitle

\tableofcontents

% List of contributors - print here or after the Bibliography.
% \PrintAtlasContribute{0.30}
\clearpage

%-------------------------------------------------------------------------------
\section{Introduction}
\label{sec:intro}

Final states consisting of a photon and a jet (\gj) with
large invariant mass could be produced in proton--proton ($pp$)
collisions at the Large Hadron Collider (LHC) in
many scenarios of physics beyond the Standard Model (SM),
including decays of excited quarks ($q^*$) and
non-thermal quantum black holes.

\begin{figure}[htb]
\begin{center}
  \includegraphics[width=0.8\textwidth]{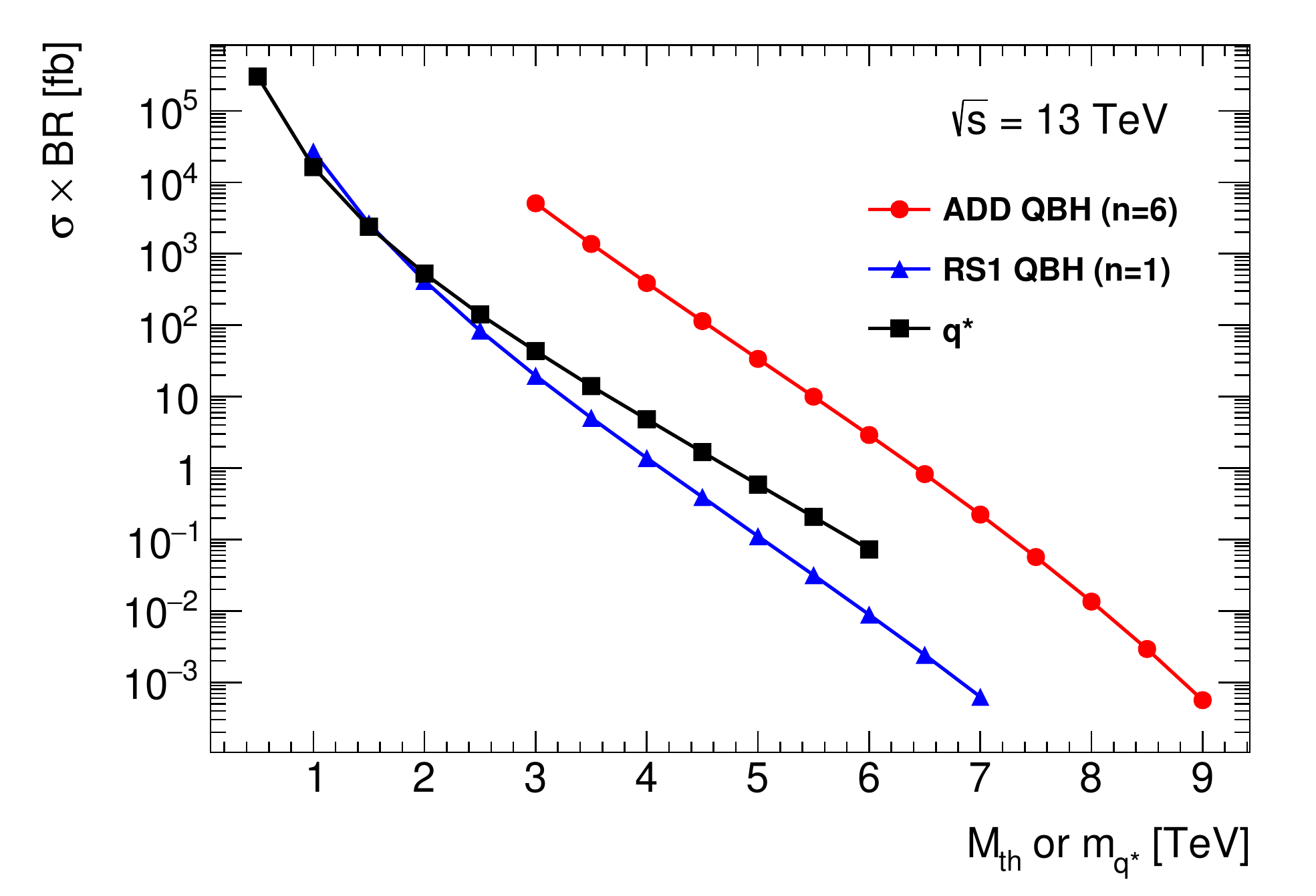}
  \caption{Production cross section times $\gj$ branching ratio
    for an excited quark $q^*$
    and two different non-thermal quantum black
    hole models (RS1, ADD) as a function of the $q^*$ mass
    or the mass threshold for black hole production $M_\mathrm{th}$,
    in $pp$ collisions at $\sqrt{s}=13$~TeV.
    The $q^*$ cross section is computed at leading order in $\alpha_{\rm s}$
    with the \pythia~8.186 event generator~\cite{pythia8}.
    The excited-quark model assumes that the compositeness scale
    is equal to the excited-quark mass $m_{q^*}$, and that
    gauge interactions of excited quarks are like those in the SM,
    with the $SU(3)$, $SU(2)$, and $U(1)$ coupling
    multipliers fixed to $f_{\rm s} = f = f' = 1$.
    The quantum black hole cross sections are obtained with the
    QBH~2.02 event generator~\cite{Gingrich:2009da}.
    In the RS1 and ADD quantum black hole models the number of
    extra spatial dimensions is $n=1$ and $n=6$, respectively.
    The maximum mass for black hole production is set to the $pp$
    centre-of-mass energy or to $3M_\mathrm{th}$ if $M_\mathrm{th}<\sqrt{s}/3$.
    The cross sections are calculated in 0.5 GeV mass steps
    (dots) and interpolated with a continuos function (solid lines).
    \label{fig:xsec_qstar_and_qbh}}
\end{center}
\end{figure}

Excited-quark states with vector-like couplings to the SM
particles~\cite{Bhattacharya:2009p2168,PhysRevD42815}
may be produced in $pp$ collisions via the fusion of a gluon
with a quark and then decay promptly to a quark and a photon
($qg\to q^{*} \to q\gamma$).
At a $pp$ centre-of-mass energy of $\sqrt{s}=13$~TeV,
the expected leading-order (LO) $q^*$ production cross sections ($pp\to q^* +X$)
times the $q^*\to q\gamma$ decay branching ratios,
combining all flavours of excited quarks and assuming a
compositeness scale equal to the excited-quark mass $m_{q^*}$,
are shown in figure~\ref{fig:xsec_qstar_and_qbh} as a function of $m_{q^*}$.
These cross sections were obtained with the \pythia~8.186
event generator~\cite{pythia8}.
Only gauge interactions like those in the SM are considered for the
excited quarks, with the $SU(3)$, $SU(2)$, and $U(1)$ coupling
multipliers fixed to $f_\mathrm{s} = f = f^\prime = 1$.
The predicted production cross section times branching ratio is
approximately 5~fb for $m_{q^{*}} = 4$~TeV.

Theories with $n$ extra spatial dimensions, such as
the Randall--Sundrum type-I (RS1) model~\cite{Randall:1999ee}
and the Arkani-Hamed--Dimopoulous--Dvali (ADD)
model~\cite{ArkaniHamed:1998rs,Antoniadis:1998ig},
solve the mass hierarchy problem of the SM by lowering the fundamental
scale $M_*$ of quantum gravity ($\tilde M$ in the RS1 model and
$M_{\rm D}$ in the ADD model) to a few~\tev.
As a consequence, the LHC could produce quantum black holes (QBH) with
masses near $M_*$~\cite{Dimopoulos:2001hw,Giddings:2001bu}, which
would then decay before thermalizing, producing
low-multiplicity final states~\cite{Calmet:2008dg,Gingrich:2009hj}.
The RS1 model studied in this article has $n = 1$ extra dimensions.
For the ADD model, the same benchmark scenario ($n = 6$)
is investigated as in the previous ATLAS publication~\cite{EXOT-2012-22}.
In this article it is also assumed that the mass threshold for black hole production
is equal to the Planck scale, $M_\mathrm{th} = M_*$.
The maximum mass for black hole production, which in any case cannot exceed
the $pp$ centre-of-mass energy, is set
to $3M_*$ (when $M_*<\sqrt{s}/3$), to avoid the high-mass
regime in which a classical description of the black hole
should replace the quantum one.
A continuum of black holes with invariant masses between the threshold mass
and the maximum mass can therefore be produced, with a probability rapidly
decreasing with the mass.
The total expected production cross sections times decay branching
ratios for $pp\to \textrm{QBH} + X\to \gamma + q/g + X$
as a function of the threshold mass,
assuming that all QBHs decay to two-body final states and
summing over all parton types in the initial and final state,
are shown in figure~\ref{fig:xsec_qstar_and_qbh}.
These cross sections were obtained with the
QBH~2.02 event generator~\cite{Gingrich:2009da}.
At $\sqrt{s}=13$~TeV the predicted total production cross section
times branching ratio is 1.4~fb (390~fb) for RS1 (ADD) black holes with $M_\mathrm{th} = 4$~TeV.
%for RS1 black holes with $M_\mathrm{th} = 4$~TeV is about 1.4~fb,
%while for ADD black holes with $M_\mathrm{th} = 6$~TeV it is approximately 3~fb.

Both the $q^*\to q\gamma$ and $\mathrm{QBH}\to q\gamma,g\gamma$ decays
(regardless of the number of extra dimensions)
would yield final states with a photon and a jet having large transverse momenta
and large invariant mass $m_{\gamma j}$.
Such events would manifest themselves in the $m_{\gamma j}$ distribution as a
broad peak above the steeply falling background
from SM prompt \gj events~\cite{ATLAS:2012ar,Aad:2013gaa},
typically produced by QCD Compton scattering ($qg\to q\gamma$).

In this article, a search for a localized, high-mass
excess in the \gj invariant mass distribution is presented.
The excess would arise from $s-$channel production
of a resonant signal.
The measurement uses \datalumi\ of $pp$ collisions
collected at a centre-of-mass energy $\sqrt{s}=13$~TeV by the ATLAS
detector in 2015.

The results are interpreted in terms of the visible
cross section (i.e. the product of the production
cross section, the branching ratio, the detector acceptance and
the selection efficiency) of a generic Gaussian-shaped signal with
mass $M_{\mathrm{G}}$ and width $\sigma_{\mathrm{G}}$.
The results are also interpreted in terms of
the cross section times branching ratio to a photon
and a quark or a gluon in three benchmark models: a $q^*$ state,
a non-thermal RS1 QBH, and a non-thermal ADD QBH (for $n=6$).

For the case of a Gaussian-shaped signal, the width $\sigma_{\mathrm{G}}$ is
assumed to be proportional to $M_{\mathrm{G}}$; three possible values of
$\sigma_{\mathrm{G}}/M_{\mathrm{G}}$ are considered: 2\%, 7\% and 15\%.
The experimental photon+jet invariant mass resolution improves
from 2.4\% at 1~TeV to 1.5\% at 6~TeV.
The smallest value of $\sigma_{\mathrm{G}}/M_{\mathrm{G}}$ (2\%) thus corresponds to
the typical photon+jet invariant mass resolution and
hence represents the case of an intrinsically narrow resonance.

The RMS width of the $q^*$ lineshape is expected
to increase from about 250~GeV at $m_{q^*}=1$~TeV
to more than 1~TeV at $m_{q^*}=6$~TeV and beyond.
Quantum black holes are expected to produce even broader signals, due
to the production of a continuum of QBHs with masses between the
threshold mass $M_\mathrm{th}$ and the maximum mass.

Previous searches for generic Gaussian-shaped resonances, excited quarks,
and ADD quantum black holes in the \gj final state have been performed by the
ATLAS and CMS collaborations using $pp$ collisions at either
$\sqrt{s}=7$~TeV~\cite{ATLAS:2011ai} or 8~TeV~\cite{EXOT-2012-22,Khachatryan:2014aka}.
No significant excess of events over the background was found,
leading to lower limits on the mass
of excited quarks at 3.5~TeV from both the ATLAS and
CMS experiments~\cite{EXOT-2012-22,Khachatryan:2014aka} and on the ADD QBH
mass at 4.6~TeV by ATLAS (for $n=6$)~\cite{EXOT-2012-22}.
No limits on RS1 quantum black holes with the photon+jet final
state have been set so far.
Using the dijet final state, ATLAS has set a lower limit on the $q^*$
mass at 4.1~TeV and on the ADD QBH mass at 5.8~TeV for
$n=6$~\cite{Aad:2014aqa},
while CMS has excluded $q^*$ masses below 3.3~TeV and ADD QBH masses
below 4.0--5.3~TeV depending on $n$~\cite{CMS:2012yf,Chatrchyan:2013qha}.
ATLAS has also searched for quantum black hole production by looking
for high-mass dilepton~\cite{Aad:2014cka} and lepton+jet~\cite{Aad:2013gma}
resonances in $\sqrt{s}=8$~TeV data.

Recently, using data collected at $\sqrt{s} = \SI{13}{\TeV}$,
the ATLAS and CMS collaborations performed searches
for excited quarks in dijet final states~\cite{ATLAS:2015nsi, Khachatryan:2015dcf},
and ATLAS searched for quantum black holes in dijet final states~\cite{ATLAS:2015nsi} and
for thermal black holes in multijet final states~\cite{Aad:2015mzg}.
Excited quarks with masses below 5.2~TeV
and RS1 or ADD ($n=6$) quantum black holes with masses below
5.3~TeV and 8.3~TeV respectively, decaying to dijets, are excluded.

The searches presented in this article exploit analysis
techniques and a selection strategy similar to those in a previous
search using 20.3~\ifb\ of $pp$ collisions
at $\sqrt{s}=8$~TeV~\cite{EXOT-2012-22}.
Despite the six times smaller integrated luminosity at
$\sqrt{s}=13$~TeV, the sensitivity of the present search to
the $q^*$ and QBH signals exceeds the exclusion limits
obtained with 8~TeV data.
This is due to the significant growth of the
$q^*$ and QBH production cross section with the increase
of the $pp$ centre-of-mass
energy from 8~TeV to 13~TeV. For instance, for a mass of 5~TeV,
the production cross sections rise by more than two orders of
magnitude for both the $q^*$ and QBHs, while the background cross
section increases by less than an order of magnitude.

The article is organized as follows.
In section~\ref{sec:detector} a brief description of the ATLAS detector
is given.
Section~\ref{sec:samples} summarizes the data and simulation samples
used in this study.
The event selection is discussed in section~\ref{sec:selection}.
The signal and background modelling are presented in section~\ref{sec:models}.
The systematic uncertainties are described in section~\ref{sec:systematics}.
In section~\ref{sec:limits}, the signal search and limit-setting
strategies are discussed, and finally the results are presented in
section~\ref{sec:results}.

%-------------------------------------------------------------------------------

%-------------------------------------------------------------------------------
\section{The ATLAS Detector}
\label{sec:detector}

The ATLAS detector~\cite{PERF-2007-01} is a multi-purpose particle
detector with approximately forward-backward symmetric
cylindrical geometry.\footnote{ATLAS uses a right-handed coordinate
  system with its origin at the nominal interaction point (IP) in the
  centre of the detector and the $z$-axis along the beam pipe. The
  $x$-axis points from the IP to the centre of the LHC ring, and the
  $y$-axis points upward. Cylindrical coordinates $(r,\phi)$ are used
  in the transverse plane, $\phi$ being the azimuthal angle around the
  $z$-axis. The pseudorapidity is defined in terms of the polar angle
  $\theta$ as $\eta=-\ln\tan(\theta/2)$.}
The inner tracking detector (ID) covers $\left| \eta \right|<2.5$ and
consists of a silicon pixel detector (including the newly installed
innermost pixel layer~\cite{Capeans:1291633}), a silicon microstrip detector,
and a straw-tube transition radiation tracker.
The ID is surrounded by a thin
superconducting solenoid providing a 2~T axial magnetic field
and by a high-granularity lead/liquid-argon (LAr) sampling
electromagnetic (EM) calorimeter. The EM
calorimeter measures the energy and the position of
electromagnetic showers with $\left| \eta \right|<3.2$.
It includes a presampler (for $|\eta|<1.8$) and three sampling layers,
longitudinal in shower depth, up to $|\eta|=2.5$.
The hadronic calorimeter, surrounding the electromagnetic one and
covering $\left| \eta \right|<4.9$, is a sampling calorimeter
which uses either scintillator tiles or LAr as the active medium, 
and steel, copper or tungsten as the absorber
material.
The muon spectrometer (MS) surrounds the calorimeters and consists of three
large superconducting air-core toroid magnets, each with eight coils,
a system of precision tracking chambers ($\left| \eta \right|<2.7$),
and fast tracking chambers ($\left| \eta \right|<2.4$) for triggering.

Events containing photon candidates are selected by a two-level
trigger system.
The first-level trigger is hardware based; using a trigger cell
granularity coarser than that of the EM
calorimeter, it searches for electromagnetic clusters within a fixed
window of size $0.2\times 0.2$ in $\eta\times\phi$ and retains only
those whose total transverse energy in two adjacent trigger cells is
above a programmable threshold.
The second, high-level trigger is implemented in software and
employs algorithms similar to those used offline to identify jets
and photon candidates.
Such algorithms exploit the full granularity and precision of the
calorimeter to refine the first-level trigger selection, based on
the improved energy resolution and detailed information about energy
deposition in the calorimeter cells.

%-------------------------------------------------------------------------------

%-------------------------------------------------------------------------------
\section{Data and simulation samples}
\label{sec:samples}

Data events were collected in $pp$ collisions at $\sqrt{s}=13$~TeV
produced by the LHC in 2015.
The average number of inelastic interactions per bunch crossing was 14.
Only events taken in stable beam conditions and in which the trigger system,
the tracking devices and the calorimeters were operational and with good data
quality are considered. The integrated luminosity of the analysed data sample is
$L_\mathrm{int}=$~\datalumi.

The events used for the analysis are recorded by a trigger requiring
at least one photon candidate with transverse momentum above
140~GeV and passing loose identification requirements based on
the shower shapes in the EM calorimeter and on the
energy leaking into the hadronic calorimeter from the EM calorimeter~\cite{Aad:2010sp}.

Twelve samples of simulated $pp\to q^* + X\to \gamma + q + X$ events,
with $q^*$ masses in the range between 500~GeV and 6~TeV
and separated by 500~GeV intervals,
were generated at leading order in the strong coupling
constant $\alpha_{\rm s}$ with \pythia~8.186.
The NNPDF 2.3~\cite{Ball:2012cx} parton distribution functions and
the A14 set of tuned parameters~\cite{uetune} of the underlying
event were used.

Simulated samples of QBHs decaying into a photon and a quark or a gluon
were generated with QBH 2.02,
interfaced to \pythia~8.186 for hadronization and simulation of the
underlying event.
The CTEQ6L1~\cite{cteq6} parton distribution functions and the A14
tune of the underlying event were used.
Thirteen samples of $pp\to \mathrm{QBH} + X\to \gamma + q/g +X$ events
were produced for RS1 (ADD $n = 6$) quantum black holes with equally spaced
$M_\mathrm{th}$ values between 1 (3)~TeV and 7 (9)~TeV, in 0.5~TeV
steps.

To study the properties of the background,
events from SM processes containing a photon with associated jets
are simulated using the \sherpa 2.1.1~\cite{Gleisberg:2008ta}
generator, requiring a photon transverse momentum
above 70~GeV. Matrix elements are calculated at LO with up to
four partons and merged with the \sherpa parton
shower~\cite{Schumann:2007mg} using the ME+PS@LO
prescription~\cite{Hoeche:2009rj}. The CT10 PDF
set~\cite{cteq10} is used in conjunction with a dedicated
parton shower tuning developed by the \sherpa authors.
The samples are binned in the photon transverse
momentum, $p_\rm{T}^\gamma$, to cover the full
spectrum relevant to this analysis.

All the above Monte Carlo (MC) samples were passed through a detailed
\textsc{Geant4}~\cite{Agostinelli2003250} simulation of the ATLAS detector
response~\cite{Aad:2010ah}. Moreover, additional inelastic
$pp$ interactions in the same and neighbouring bunch crossings,
denoted as pile-up, are included in the event simulation
by overlaying a number of minimum-bias events consistent with that
observed in data.
Multiple overlaid proton--proton collisions are simulated
with the soft QCD processes of \pythia 8.186 using the A2 tune~\cite{Atlas:ATL-PHYS-PUB-2012-003,}
and the MSTW2008LO PDF set~\cite{Martin:2009iq}.

Supplementary studies of the invariant mass shape of the \gj
background are also performed with the parton-level,
next-to-leading-order (NLO) \jetphox v1.3.1\_2 generator~\cite{Catani:2002ny}
using the NNPDF 2.3 parton distribution functions
and the NLO photon fragmentation function~\cite{Bourhis2001}.
The nominal renormalization, factorization and
fragmentation scales are set to
the photon transverse momentum.
Jets of partons are reconstructed using
the anti-$k_t$ algorithm~\cite{Cacciari:2006}
with a radius parameter $R = 0.4$.
The total transverse energy from partons produced inside a
cone of size $\Delta R =
\sqrt{(\Delta\eta)^2+(\Delta\phi)^2} = 0.4$ around the photon
is required to be lower than $2.45~\mathrm{GeV}+0.022\times p_\rm{T}^\gamma$,
to match the selection requirement described in the next section.
Experimental effects (detector reconstruction efficiencies and resolution)
as well as hadronization and pile-up are not taken into account in this sample.

%-------------------------------------------------------------------------------

%-------------------------------------------------------------------------------
\section{Event selection}
\label{sec:selection}

% primary vertex
Each event is required to contain at least one primary vertex candidate
with two or more tracks with $\pt > 400$~MeV.
The tracks must satisfy quality requirements based on the number of
reconstructed intersections with the silicon pixel and strip
detectors and the track impact parameters with respect to the centre
of the luminous region.
The primary vertex is defined as the candidate with the largest
sum of the $\pt^2$ of the tracks that are considered to be associated
to it, based on a requirement on a $\chi^2$ variable calculated
between the estimated vertex position and the point of closest
approach of the track to the vertex.

% photon reconstruction and selection
Photons are reconstructed from energy deposits (clusters) found
in the EM calorimeter by a sliding-window algorithm.
The reconstruction algorithm looks for matches between energy
clusters and tracks reconstructed in the inner detector and
extrapolated to the calorimeter.
Well-reconstructed clusters matched to tracks are classified as
electron candidates while clusters without matching tracks are
classified as unconverted photon candidates.
Clusters matched to pairs of tracks that are consistent with the
hypothesis of a $\gamma\to e^+e^-$ conversion process are classified
as converted photon candidates.
To maximize the reconstruction efficiency for electrons and photons,
clusters matched to single tracks without hits in an active region of
the innermost pixel layer are considered as electron candidates and
as converted photon candidates. Both unconverted and converted photon
candidate are used for the search presented in this paper.

The energies of the photon candidates are calibrated
following the procedure described in ref.~\cite{Aad:2014nim}.
The calibration algorithm, tuned using 13~TeV event simulation,
accounts for energy loss upstream of the EM calorimeter and for both
lateral and longitudinal shower leakage.
Correction factors are extracted from 8~TeV $Z\to ee$ data and
simulated events reconstructed with the algorithms used in the 2015
data taking.
Additional corrections and systematic uncertainties take into account
the differences between the 2012 and 2015 configurations.

To reduce backgrounds from hadrons, photon candidates are required
to fulfil $\eta$-dependent requirements consisting of
nine independent selections, one on the hadronic leakage and eight on
shower shape variables measured with the first two
sampling layers of the electromagnetic calorimeter~\cite{Aad:2010sp}.
The requirements were optimized for the 2015 data-taking conditions
using simulated samples of photons and hadronic jets produced in
13~TeV $pp$ collisions.
The simulation is corrected for the differences between
$\sqrt{s}=8$~TeV data and simulated events for each photon shower shape
variable.
%% and attributed to an imperfect modelling of the detector
%% material and of the evolution of the photon electromagnetic shower
%% in the calorimeter.

Groups of contiguous calorimeter cells (topological clusters)
are formed based on the significance of the ratio of deposited energy
to calorimeter noise.
To further reduce background photons from hadronic jets, the transverse
isolation energy \etiso of the photon candidates is required to be
%% less than $2.45~\mathrm{GeV} + 0.022\times \etgamma$.
less than $2.45~\mathrm{GeV} + 0.022\times p_{\rm T}^\gamma$.
This energy is computed from the sum of the energies of all
cells belonging to topological clusters and within a cone of
$\Delta R$ = 0.4 around the photon direction.
The contributions from the underlying event and the
pile-up~\cite{Cacciari:area,Cacciari:UE}, as well as
from the photon itself, are subtracted.
The isolation requirement has a signal efficiency of about 98\% over
the whole photon transverse momentum range relevant to this analysis.

% jet selection
Jets are reconstructed from topological clusters of calorimeter
cells using the anti-$k_t$ algorithm with radius parameter $R=0.4$.
Jets affected by noise or hardware problems in the detector, or
identified as arising from non-collision backgrounds, are
discarded~\cite{ATLAS:2016323}.
Jet four-momenta are computed by summing over the topological
clusters that constitute each jet, treating each
cluster as a four-vector with zero mass.
To reduce the effects of pile-up on the jet momentum, an area-based
subtraction method is employed~\cite{Cacciari:area,Cacciari:UE}.
Jet energies are then calibrated by using
corrections from the simulation and scale factors determined in
various control samples (\gj, $Z$+jet and multijet events) in 8~TeV
data~\cite{Aad:2014bia} and validated with early 2015
data~\cite{ATL-PHYS-PUB-2015-015}.
These corrections are applied to 2015 data, after
taking into account in the simulation the
changes in the detector and in the data-taking conditions
between 8~TeV and 13~TeV data,
and propagating as systematic uncertainties those related
to this extrapolation procedure.
Jets with $\pT<20$~GeV or within $\Delta R = 0.2$ (0.4)
of a well-identified and isolated electron (photon) with
transverse momentum above 25~GeV are not considered.

%
% event selection
%
%
Events are selected if they contain at least one photon candidate and
at least one jet candidate satisfying all the previous criteria and each
having $\pt > 150$~GeV.
The photon trigger has an efficiency of $\left(99.9^{+0.1}_{-1.3}\right)\%$
for these events.
The trigger efficiency is measured in data as the product of the efficiency of
the high-level trigger computed from events selected by the first-level
trigger and the efficiency of the first-level trigger with respect
to offline identification~\cite{ATLAS-CONF-2012-048}.

Since $t$-channel \gj and dijet production rates increase
while the rate of the $s$-channel signal production decreases with the
photon and jet absolute pseudorapidity, photons are required to be in the
barrel calorimeter, $|\eta^\gamma| < 1.37$.
Moreover, as a consequence of the different production mechanisms
for the signal and background, the pseudorapidity
separation $\Delta \eta$ between the photon and the jet candidates
tend to be smaller for the signal than for the background, particularly
for large values of the photon--jet invariant mass.
For this reason, events with $|\Delta \eta|>1.6$ are discarded.

In events in which more than one good photon or jet candidate is found,
the highest-\pT candidate of each type is selected to form the resonant
\gj candidate.
Events in which the angular separation between the photon and any
jet with $\pT>30$~GeV (after the jet--photon overlap removal) is
$\Delta R<0.8$ are discarded. This requirement suppresses background
events from SM photon+jet production in which the photon is emitted
at large angles in the fragmentation of a quark or a gluon.

The total signal efficiency (including detector acceptance) depends
on the resonant mass of the hypothetical signal,
and is described in the next section.
The product of acceptance times efficiency for the
QBH and $q^*$ signals and a resonance mass of 3~TeV is
close to 50\%.

There are 2603 candidates in the data sample passing the full event
selection and having invariant mass above 1 TeV.
The \gj candidate with the highest \mgj\ value has an
invariant mass of 2.87~TeV.
A small contamination from dijet events
is expected, due to jets misidentified as photons in the calorimeter.
Such fake candidates typically arise from jets containing a neutral
meson (most likely a $\pi^0$) carrying a large fraction of the jet
energy and decaying into two collimated photons.
These candidates are on average less isolated from activity
in the neighbouring cells of the calorimeters and have on average
wider shower shapes in the electromagnetic calorimeter.
The purity of true photon+jet events in the selected sample is estimated
to be around 93\% by means of a two-dimensional sideband method
based on the numbers of photon+jet candidates in which the photon either passes
anti-isolation requirements, anti-identification requirements, or both~\cite{Aad:2010sp},
thus indicating that the dijet contamination is rather small.
For the purity measurement a photon candidate is considered
non-isolated if its isolation transverse energy is at least 3~GeV
larger than the maximum allowed $\etiso$ for a photon to be regarded
as isolated.
A photon candidate is non-identified if it fails at least one of the
requirements on four shower shape variables computed from the energy
deposited in the finely segmented cells of the first layer of the
electromagnetic calorimeter.
No evidence of a dependence of the photon purity on the \gj invariant mass is
observed.

Figure~\ref{fig:comparison_data_sherpa_jetphox} shows the
comparison between the \mgj\ distribution of events selected in data
and the shapes predicted by \sherpa and \jetphox for SM \gj
production, neglecting the dijet contribution.
The bin width rises from 100~GeV to 300~GeV with increasing
\mgj, to account for the corresponding decrease in the number of data events and increase in
the intrinsic width of most of the signals considered in this study.
The simulated spectra from \sherpa and \jetphox are normalized
to the data in the range 0.5~TeV$<\mgj<$2.5~TeV.
The shapes of the \mgj\ distributions in data and simulation agree in the range studied.
The agreement is
better for \sherpa, due to the inclusion of
hadronization and underlying-event effects and of the detector response.
\begin{figure}[htbp]
  \centering
  \includegraphics[width=0.8\textwidth]{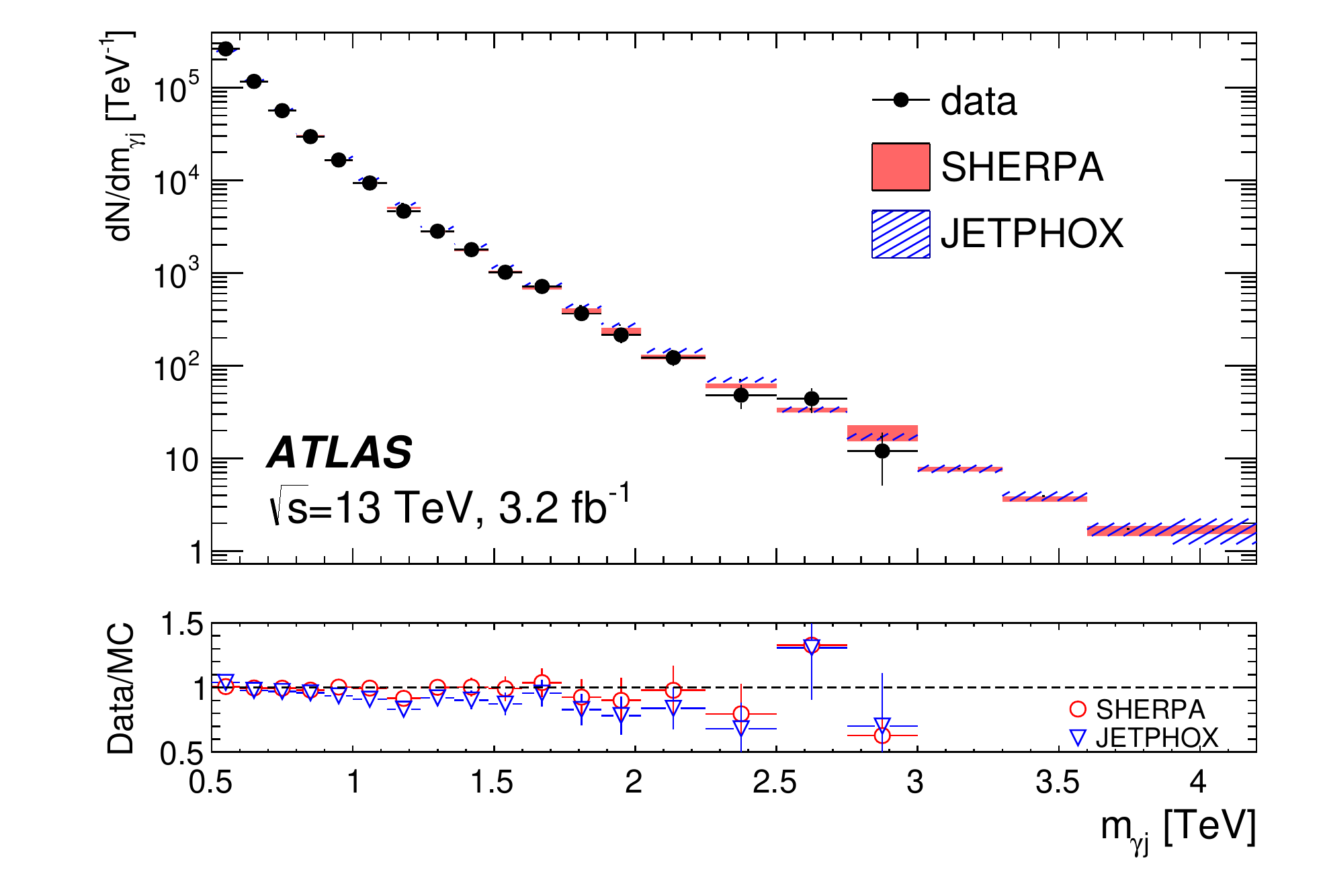}
  \caption{Photon--jet invariant mass spectrum in data (black dots) compared
    to the shape predicted by a $\gamma+$jet parton-level calculation
    (\jetphox, hatched blue bands) and a parton-shower simulation
    (\sherpa, solid red bands). Both predictions are normalized to
    data in the range $0.5~\TeV<\mgj<2.5$~TeV.
    The number of events in each bin is normalised by the bin width.
    In the bottom panel the ratios of the data to the two
    predictions are shown. Statistical uncertainties due to the size of the 
    data and simulated samples and systematic uncertainties for simulations are shown.
  }
  \label{fig:comparison_data_sherpa_jetphox}
\end{figure}

%-------------------------------------------------------------------------------

%-------------------------------------------------------------------------------
\section{Signal and background models}
\label{sec:models}
In order to evaluate the strength of a possible contribution of a
signal originating from physics beyond the SM, an unbinned
maximum-likelihood fit of the signal+background model to the
\mgj\ distribution of the selected data events is performed. 

\subsection{Signal model}
\label{ssec:signal_model}

The signal model consists of the expected \mgj\ distribution
after reconstruction and selection
(called {\em template} in the following)
for each type of signal under study
and as a function of its hypothetical mass $M$
($M_{\mathrm{G}}$, $m_{q^*}$ or $M_\mathrm{th}$ depending on the
signal type).

The model is the product $f_\mathrm{sig}(\mgj) \times \Big(\sigma \times BR\Big)
\times \Big(A\times \varepsilon\Big) \times L_\mathrm{int}$
of a template $f_\mathrm{sig}(\mgj)$ with
the production cross section times the branching ratio to a
photon and a quark or a gluon, $\sigma \times BR$,
the expected acceptance times efficiency, $A\times \varepsilon$,
and the integrated luminosity of the sample, $L_\mathrm{int}$.

In the case of a generic Gaussian-shaped signal
with mass $M_{\mathrm{G}}$ and an arbitrary production cross section,
the template is a simple Gaussian distribution
centred at $M_{\mathrm{G}}$ with a width $\sigma_{\mathrm{G}}$ proportional
to the mass; three possible values of $\sigma_{\mathrm{G}}/M_{\mathrm{G}}$ are
considered (2\%, 7\%, 15\%).
The results are directly interpreted in terms of the visible
cross section $\sigma \times BR \times A \times \varepsilon$,
which is the parameter of interest in the maximum-likelihood fit
described in section~\ref{sec:limits}.

For the other three types of signal ($q^*$, RS1 QBH, ADD QBH),
the parameter of interest in the fit is the product $\sigma \times BR$.
The normalized \mgj\ template for the generated signals (section~\ref{sec:samples})
is obtained via smoothing with a kernel density estimation
technique~\cite{rosenblatt1956,parzen1962}.
The normalized distributions of the invariant masses of the candidates
passing the full selection in the simulated signal samples are used.
For intermediate masses where generated samples are not available,
the \mgj\ distribution is obtained by a moment-morphing
method~\cite{Baak:2014fta}.
The product of acceptance times efficiency $A\times \varepsilon$
for $q^*$ and QBH signals for each mass $M$ is obtained
through a continuous interpolation of the values
obtained from the simulation of the signal samples generated
at discrete mass points (section~\ref{sec:samples}). The
interpolating function is a third-order spline.
The acceptance times efficiency curves as a function
of $M$ are rather similar for the
three models, increasing between $M=1$~TeV
($A\times\varepsilon\approx 46\%$) and $M=4$ TeV
($A\times\varepsilon\approx 51\%$), beyond which it
slowly decreases ($A\times\varepsilon\approx 47\%$ at 9~TeV).

\subsection{Background model}
\label{ssec:bkg_model}

The background \mgj\ template is the same four-parameter ansatz
function~\cite{Harris:2011bh} as used in previous searches for
high-mass resonances in the \gj final state~\cite{ATLAS:2011ai,EXOT-2012-22}:
\begin{equation}
  f_\mathrm{bkg}( x \equiv \mgj / \sqrt{s}) = p_{0}(1-x)^{p_1} x^{-p_2-p_3\log{x}}.
\end{equation}
The parameters of this empirical function, as well as the total
background yield, are directly extracted from the final fit to the data
with the signal+background model.

The possible bias in the fitted signal due to choosing
this functional form is estimated through signal+background fits
to large \gj background samples generated with \jetphox,
and included in the systematic uncertainties, as described in
section~\ref{sec:systematics}.
The small contamination from dijet events is neglected, since
the photon--jet purity of the sample is high (around 93\%) and
does not depend significantly on \mgj.
Detector effects are considered by reweighting the \jetphox sample with
corrections obtained from the \sherpa photon--jet simulation, as explained
in the next section.

The range for the fit is chosen in order to have a large efficiency
for the signal and to provide a sufficiently wide mass sideband.
The mass sideband should be wide enough to determine from the data
the parameters of the background model with good precision, but not too wide
in order to suppress as much as possible any bias in the signal.
The bias is considered acceptable if it is less than 10\% of the
expected signal yield or less than 20\% of the expected
statistical uncertainty of the background.
The chosen ranges are 1~TeV$<\mgj<5.5$~TeV in the searches for
generic Gaussian-shaped resonances, $q^*$, or RS1 QBH,
and 2~TeV$<\mgj<8$~TeV in the ADD QBH search.
These ranges probe signals with masses between
1.5 and 5 TeV (Gaussian-shaped resonances, $q^*$, RS1 QBH),
or between 3 and 7 TeV (ADD QBH).

A further test to check whether the chosen ansatz function
accurately describes the expected background distribution
is performed by fitting pseudo-data generated from the
simulated \sherpa\ photon+jet events with alternative functions with
more degrees of freedom. An $F$-test is then performed to compare
the $\chi^2$ used to estimate the goodness of the nominal fit to
the $\chi^2$ of the alternative fit.
No significant decrease of the $\chi^2$ was observed
when adding more degrees of freedom to the ansatz
function used as the nominal background model.

%-------------------------------------------------------------------------------

%-------------------------------------------------------------------------------
\section{Systematic uncertainties}
\label{sec:systematics}

The systematic uncertainty of the integrated luminosity is $\pm 5\%$.
It is derived, following a methodology similar to that detailed in
ref.~\cite{Aad:2013ucp}, from a preliminary calibration of the
luminosity scale using $x$--$y$ beam-separation scans performed
in August 2015.

For the $q^*$ and QBH signals, the acceptance times efficiency
is subject to systematic uncertainties in the
trigger efficiency ($^{+0.1}_{-1.3}\%$), photon identification
efficiency ($\pm 2\%$ to $\pm4\%$) and photon isolation efficiency ($\pm 1\%$).
The systematic uncertainty on the trigger efficiency is estimated
as the difference between the efficiency measured in data
and the efficiency obtained in MC simulations.
It is dominated by the statistical uncertainty of the
measurement in data.
The photon identification and isolation efficiency uncertainties
are estimated conservatively by recomputing the signal efficiency after removing
the MC-to-data corrections from the shower shape variables and the
transverse isolation energy.

Additional uncertainties in the signal arise from the
uncertainties in the photon and jet energy scales and resolutions as a
consequence of the requirements placed on the photon and jet
transverse momenta and invariant mass.
The energy scale and resolution uncertainties have effects on $A\times\epsilon$ that are smaller than
$\pm 0.5\%$ for photons (and are thus
neglected) and are about $\pm 1\%$ to $\pm 2\%$ (energy scale) and
$\pm 1\%$ (energy resolution) for jets.
These uncertainties have a negligible impact on the signal \mgj\ distribution since
the intrinsic width dominates over the experimental
resolution and is much larger than the possible bias arising from the
photon and jet energy scale uncertainties.
The limited size of the simulated signal samples yields an
uncertainty in the signal efficiency of $\pm 1\%$.
Systematic uncertainties in the signal acceptance and shape
due to the PDF uncertainties were examined and found to be
negligible compared to the other uncertainties.

The background yield and values for the parameters of its invariant mass
distribution are directly extracted from a fit to the data.
A possible systematic uncertainty in the signal yield arises from the
choice of functional form used to model the background distribution.
In order to estimate this uncertainty, a large \gj background
sample (about seven billion events) is generated using
\jetphox\ and fit with the full signal+background model. This is done for each
tested signal model and mass $M$.
The photon and jet kinematic
requirements described in section~\ref{sec:selection} are applied.
Since no signal is present in these background-only
samples, the resulting number of spurious signal events from the fit,
$N_\mathrm{spur}(M)=\sigma^\mathrm{eff}_\mathrm{spur}(M) \times L_\mathrm{int}$ for the Gaussian-shaped signal
and
$N_\mathrm{spur}(M)=\sigma^\mathrm{eff}_\mathrm{spur}(M)\times \Big(A\times \varepsilon\Big)(M) \times L_\mathrm{int}$
for the other signal models,
is taken as an estimate of the bias for the model under test.
In order to cover possible uncertainties in the \jetphox\ prediction itself,
the fit is repeated after varying each of several model parameters and estimating
reconstruction effects.
The final uncertainty is thus the largest spurious signal cross
section obtained when doing the signal+background fits to the
following background-only samples:
\begin{itemize}
\item the nominal sample generated with \jetphox;
\item the samples generated with \jetphox\ after varying the eigenvalues of
  the NNPDF 2.3 set by $\pm 1\sigma$;
\item the samples generated with \jetphox\ after varying the value of the
  strong coupling constant by $\pm 0.002$ around the nominal value of 0.018;
\item the samples generated with \jetphox\ after varying the renormalization,
  factorization and fragmentation scales between half and twice the photon
  transverse momentum;
\item the sample obtained after reweighting the \jetphox\ \mgj\ distribution
  by the ratio of the reconstructed and particle-level \mgj\ spectra
  predicted by \sherpa.

\end{itemize}

All samples are rescaled so that the total number of events is equal to the number observed in the data.
For the Gaussian signal search, the spurious cross section
$\sigma^\mathrm{eff}_\mathrm{spur}(M)$
varies between 10~fb at 1~TeV and less than 0.1~fb at 5.5~TeV.
The estimated spurious signal yield 
$N_\mathrm{spur}(M)$
varies between $3\times 10^{-3}$ and 10\% of the expected $q^*$ yield
for masses between 1.5~TeV and 5~TeV,
between $5\times 10^{-4}$ and 10\% of the expected RS1 QBH yield
for masses between 1.5~TeV and 5~TeV, and
between $5\times 10^{-5}$ and $3\times 10^{-3}$ of the expected ADD QBH
yield for masses between 3~TeV and 7~TeV.

%-------------------------------------------------------------------------------

%-------------------------------------------------------------------------------
\section{Statistical procedures of the excess search}
\label{sec:limits}

To search for an excess over the SM background in the \mgj\ distribution
in the data, quantify its significance and
set limits, the profile-likelihood-ratio method described in
ref.~\cite{stat} is used.
The extended likelihood function $\mathcal{L}$ is built from the number
$n$ of observed events, the expected event yield $N$, and the
functions $f_\mathrm{sig}$ and $f_\mathrm{bkg}$ describing the signal and
background \mgj\ distributions:
\begin{equation}
\LL = \mathrm{Pois} (n|N(\vecth)) \prod\limits_{i=1}^nf(\mgj^i,\vecth) \times G(\vecth).
\end{equation}

In this expression $f(\mgj^i,\vecth)$ is the value of the probability
density function (pdf) of the invariant mass distribution for
each candidate event $i$, $\vecth$ represents the nuisance parameters
and $G(\vecth)$ is a set of constraints on some of the nuisance parameters,
as described in the following.

The number of expected candidates $N$ is the sum of the
number of signal events, the number of background candidates
$N_{\mathrm{bkg}}$, and the spurious signal yield $N_\mathrm{spur}(M)$
fitted on background-only samples as described in the previous section.
For the Gaussian-shaped signal, $N$ is thus:
\begin{equation}
  N = \Big(\sigma\times BR\times A\times\varepsilon\Big)(M) \times L_{\mathrm{int}} + N_{\mathrm{\mathrm{bkg}}} + N_\mathrm{spur}(M) \times \thetaspur,
\end{equation}
while for the QBH and $q^*$ signals it is:
\begin{equation}
  N = \Big(\sigma\times BR\Big)(M)\times \Big(A\times\varepsilon\Big)(\vecth^\mathrm{yield},M) \times L_{\mathrm{int}}+ N_{\mathrm{\mathrm{bkg}}} + N_\mathrm{spur}(M) \times \thetaspur .
\end{equation}
Here $\vecth^\mathrm{yield}$ are the nuisance parameters that implement the
systematic uncertainties affecting the signal yields and $\thetaspur$ is
the nuisance parameter corresponding to the systematic uncertainty
from the choice of background model.

The total pdf $f(\mgj)$ is then:
\begin{equation}
f(\mgj^i)= \frac{1}{N} \left[ \left( \Big(\sigma\times BR \times A\times\varepsilon\Big)(M) + \sigma^\mathrm{eff}_{\mathrm{spur}}(M)\times \thetaspur \right)\times L_{\mathrm{int}} \times f_{\mathrm{sig}}(\mgj^i) +  N_{\mathrm{bkg}} \times f_{\mathrm{bkg}}(\mgj^i)\right]
\end{equation}
for the Gaussian-shaped signal search and:
\begin{multline}
  f(\mgj^i) = \frac{1}{N} \left[ \left( \Big(\sigma\times BR\Big)(M) + \sigma^\mathrm{eff}_{\mathrm{spur}}(M)\times \thetaspur \right)\times\Big(A\times\varepsilon\Big)(\vecth^\mathrm{yield},M) \times L_{\mathrm{int}} \times f_{\mathrm{sig}}(\mgj^i)\right.\\ + \left. N_{\mathrm{bkg}} \times f_{\mathrm{bkg}}(\mgj^i)\right]
\end{multline}
for the QBH and $q^*$ searches, where $f_{\mathrm{sig}}$  and
$f_{\mathrm{bkg}}$ are the signal and background templates, respectively.

Apart from the spurious signal, systematic uncertainties with
an estimated size $\delta$
are incorporated into the likelihood by multiplying the relevant
parameter of the statistical model by a factor
$F_\mathrm{G}(\delta,\theta)=(1+\delta\cdot\theta)$
in the case of a Gaussian or,
for cases where a negative model parameter does not make
physical sense, by $F_\mathrm{LN}(\delta,\theta)=\mathrm{e}^{\delta \theta}$
for a log-normal pdf.
In both cases the likelihood is multiplied by a constraint term
$G(\theta)$ which is a standard normal distribution for $\theta$,
centred at zero.

The significance of the signal is estimated by computing
\pzero, which is defined as the $p$-value that quantifies the
compatibility of the data with the background-only hypothesis.
Upper limits on the signal cross section times branching ratio at 95\%
confidence level (CL) are set using a modified frequentist ($CL_{s}$)
method~\cite{cls}, by identifying the value of $\sigma\times BR$ (or
$\sigma\times BR\times A\times \varepsilon$ for the Gaussian-shaped
resonance) for which $CL_s$ is equal to $0.05$.
Due to the vanishingly small size of the selected dataset and of the
expected background at masses beyond 2.8~TeV, the results are
computed using ensemble tests.

%-------------------------------------------------------------------------------

%-------------------------------------------------------------------------------
\section{Results}
\label{sec:results}

The data distributions in the \mgj\ regions used for the final fits
are shown in figure~\ref{fig:fit_data}.
The background-only fit is overlaid, together with the expected
distribution for a few signal models.
The signal+background fits performed on the data show no significant
excess of events. The smallest $p_0$ is obtained for a mass $M$
equal to 2.6~TeV and corresponds to a significance of about $1.7\sigma$.

\begin{figure}[htbp]
 \centering
 \subfloat[]{\includegraphics[width=0.6\textwidth]{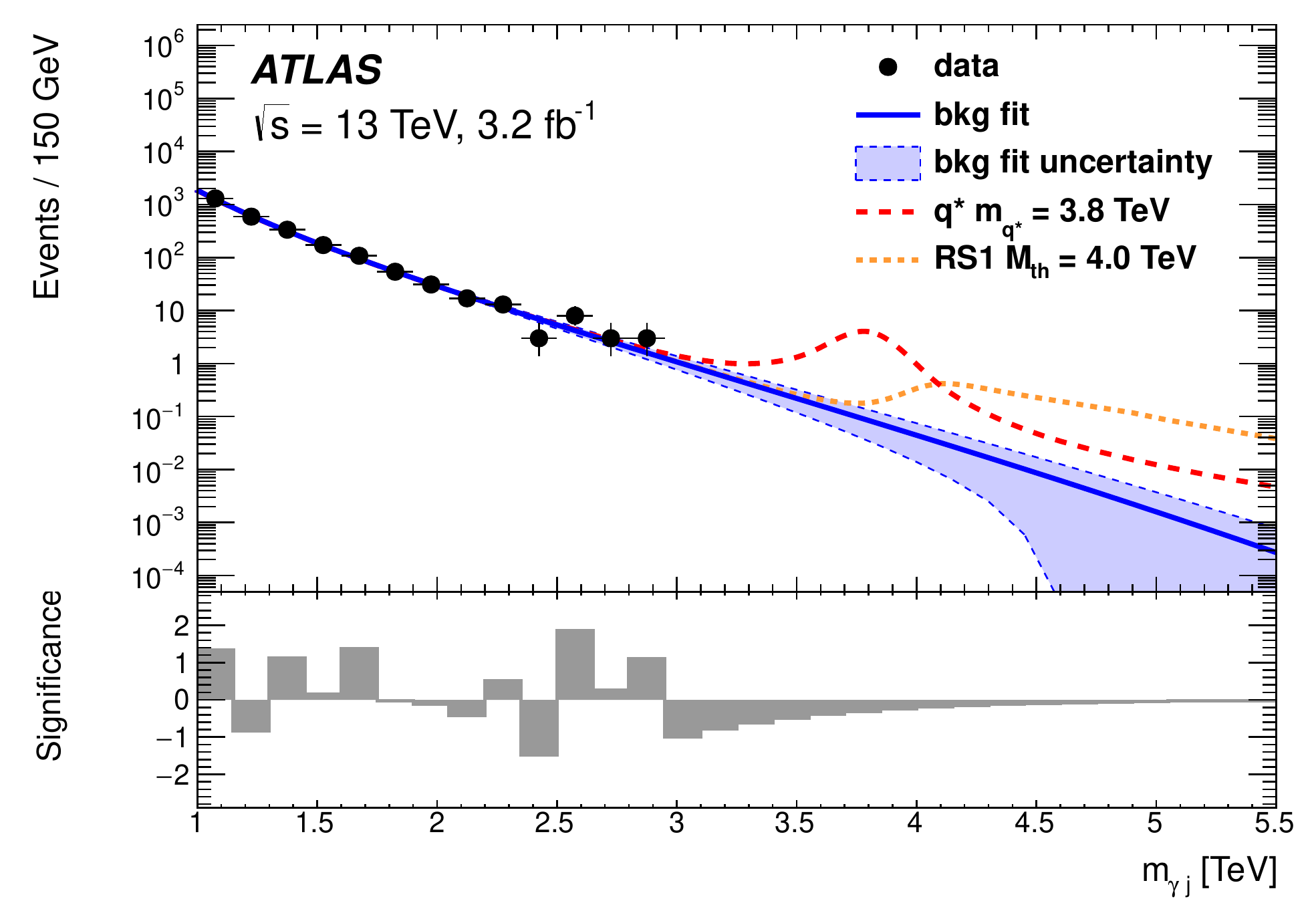}}\\
 \subfloat[]{\includegraphics[width=0.6\textwidth]{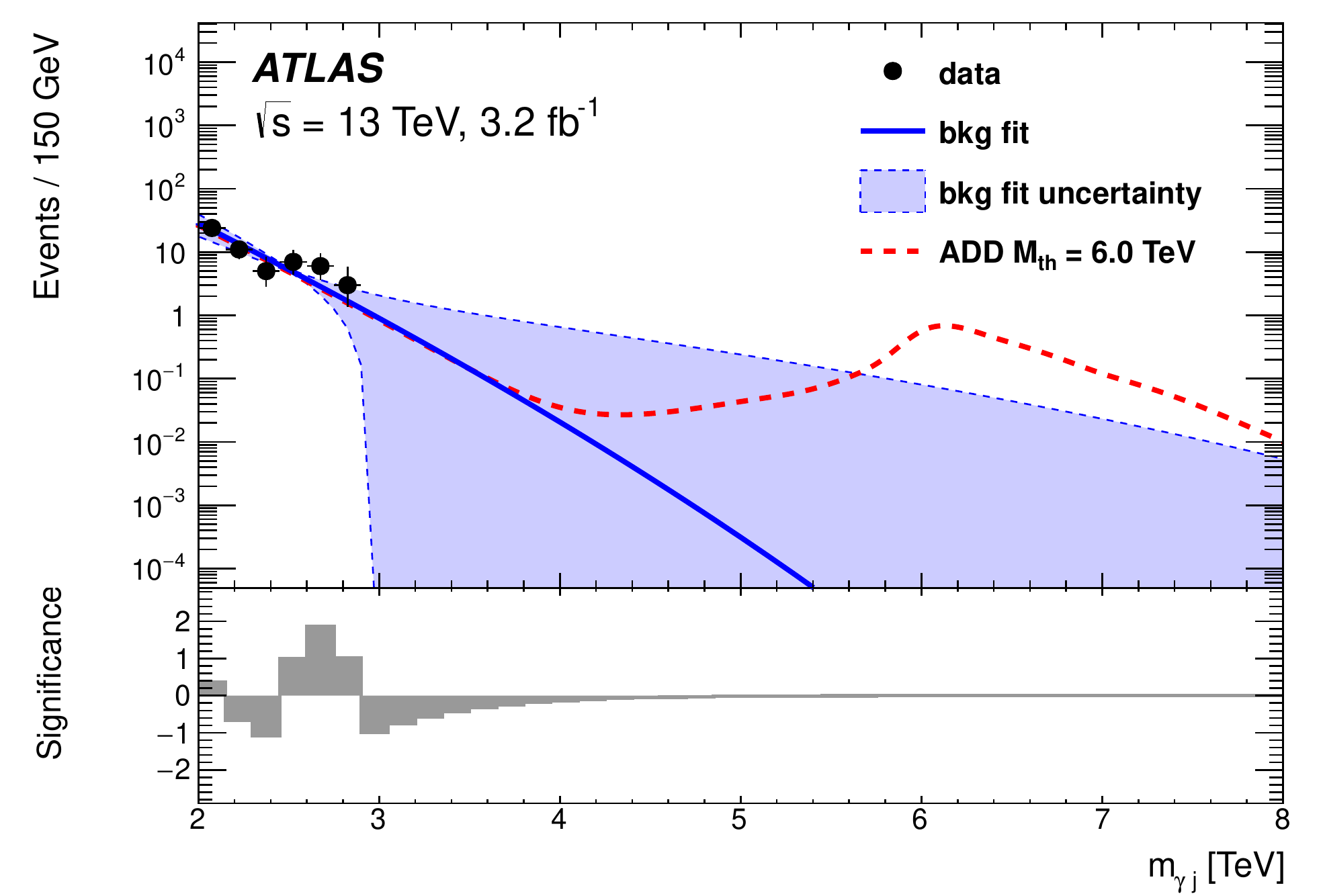}}
 \caption{Photon-jet invariant mass distributions of events
   selected in data and results of a background-only fit, for 
   (a) the $q^*$ and RS1 ($n=1$) QBH searches and (b) the 
   ADD ($n=6$) QBH search.
   The top panels show the data (dots), 
   the nominal fit results (blue lines), and the uncertainty on the
   background models (light blue bands) due to the uncertainty in the
   fit parameter values. Some examples of expected
   signals overlaid on the fitted background are also shown, for
   (a) a $q^*$ with a mass of 3.8~TeV (red dashed line) and an RS1
   ($n = 1$) QBH with a threshold mass of 4~TeV (orange dotted line)
   and (b) an ADD ($n = 6$) QBH with a threshold mass of 6~TeV (red
   dashed line).
   The bottom panels show the difference between the data and the
   prediction of the background-only fit, divided by the square
   root of the predicted background.
 }
 \label{fig:fit_data}
\end{figure}

Since no significant deviation from the background-only
hypothesis is observed, upper limits are set on the visible
cross section of a generic Gaussian-shaped signal
and on the production cross section times branching ratio
of excited quarks and quantum black holes.

The observed and expected upper limits on the visible
cross sections for a generic Gaussian-shaped signal
are shown in figure~\ref{fig:limit_data_gaussian}.
The data exclude resonances with a mass of 1.5~TeV and
visible cross sections above about 6 (50)~fb, and resonances with
a mass of 5~TeV and cross sections above about 0.8 (1.0)~fb,
for $\sigma_{\mathrm{G}}/M_{\mathrm{G}}=2\%$ (15\%).
\begin{figure}[htbp]
  \begin{center}
    \includegraphics[width=0.6\textwidth]{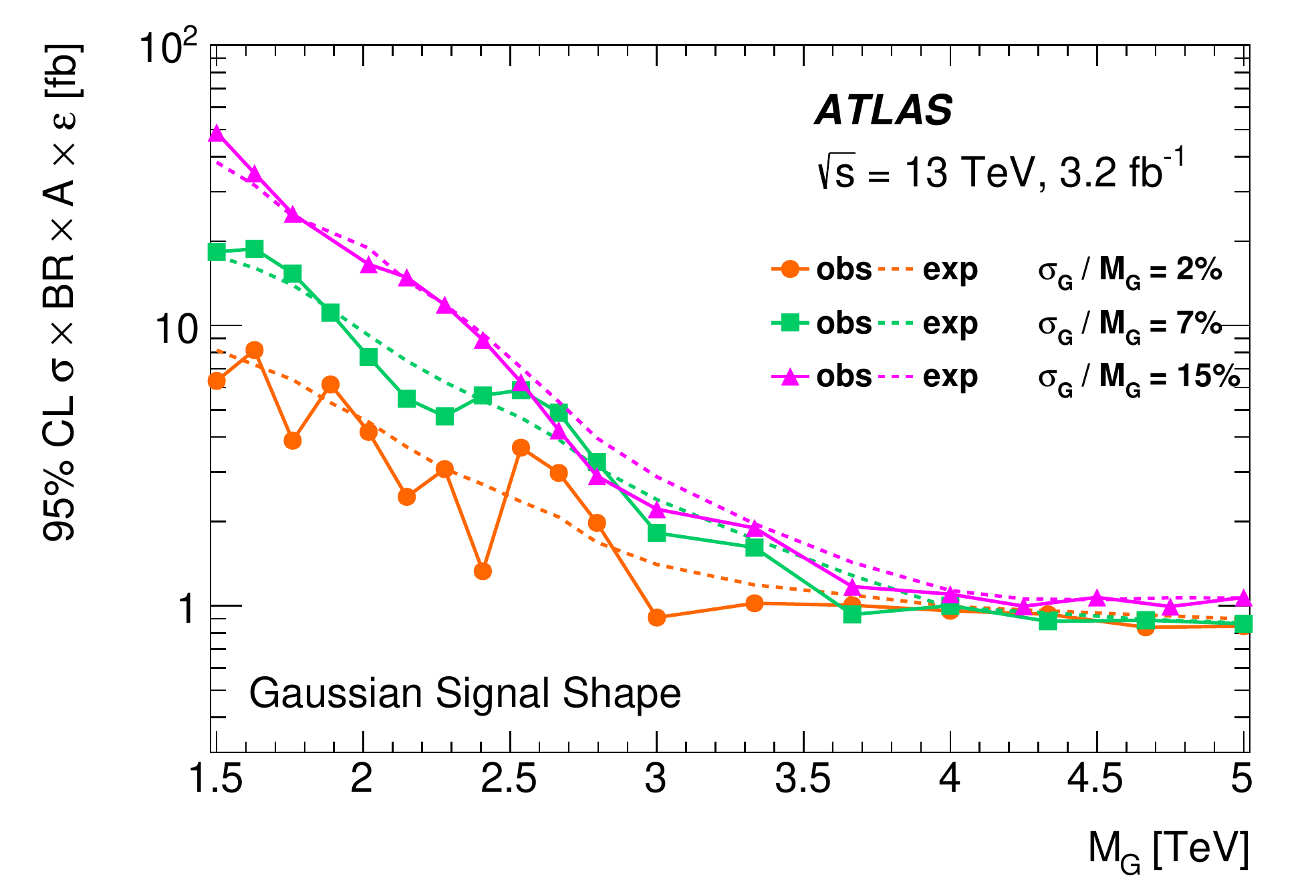}
     \caption{
       Observed (solid lines) and expected (dashed lines) 95\% CL limits on the
       visible cross section $(\sigma \times BR\times A\times \varepsilon)$
       for a hypothetical signal with a Gaussian-shaped \mgj\
       distribution as a function of the signal mass $M_{\mathrm{G}}$
       for three values of the width-to-mass ratio $\sigma_{\mathrm{G}}/M_{\mathrm{G}}$.
     }
     \label{fig:limit_data_gaussian}
  \end{center}
\end{figure}

The observed and expected upper limits on the
production cross sections times branching ratio
to a photon and a quark or a gluon
for benchmark models of excited quarks,
RS1 QBHs and ADD QBHs are shown in
figure~\ref{fig:limit_data}.

Comparing the measured upper limits
to the theoretical predictions
as a function of the mass of the resonance,
lower limits are set for the excited-quark mass at 4.4~TeV
and for the RS1 (ADD) quantum black hole mass at 3.8 (6.2)~TeV.
The uncertainty in the $q^*$ theoretical cross section
arising from PDF uncertainties reduces the maximum excluded mass by 1.5\%.
The limits on the $q^*$ and ADD QBH mass
improve on the ATLAS results at $\sqrt{s} = 8$~TeV in this channel by 0.9~TeV and 1.7~TeV, respectively.

\begin{figure}[htbp]
  \begin{center}
    \subfloat[]{\includegraphics[width=0.54\textwidth]{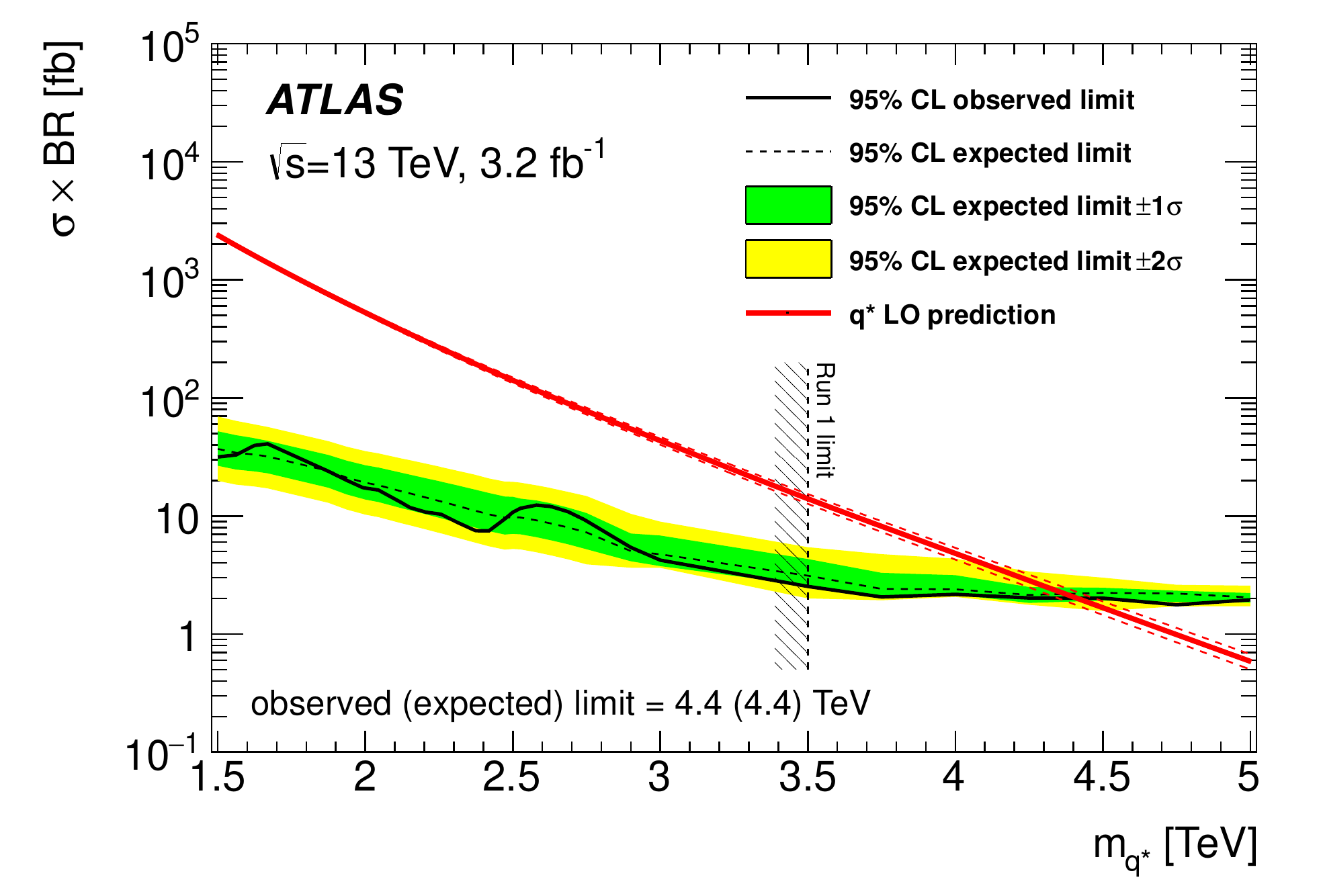}}\\
    \subfloat[]{\includegraphics[width=0.54\textwidth]{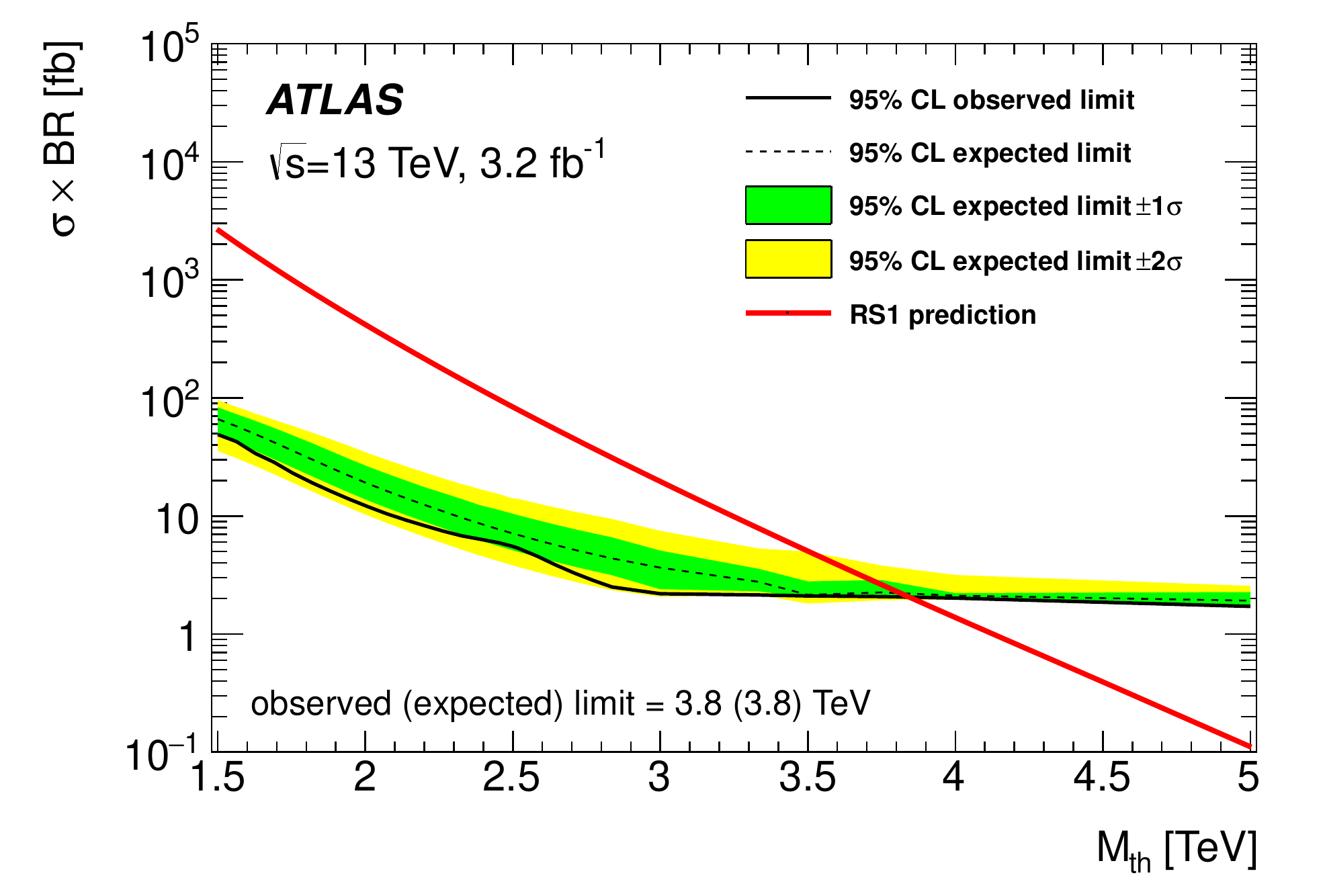}} \\
    \subfloat[]{\includegraphics[width=0.54\textwidth]{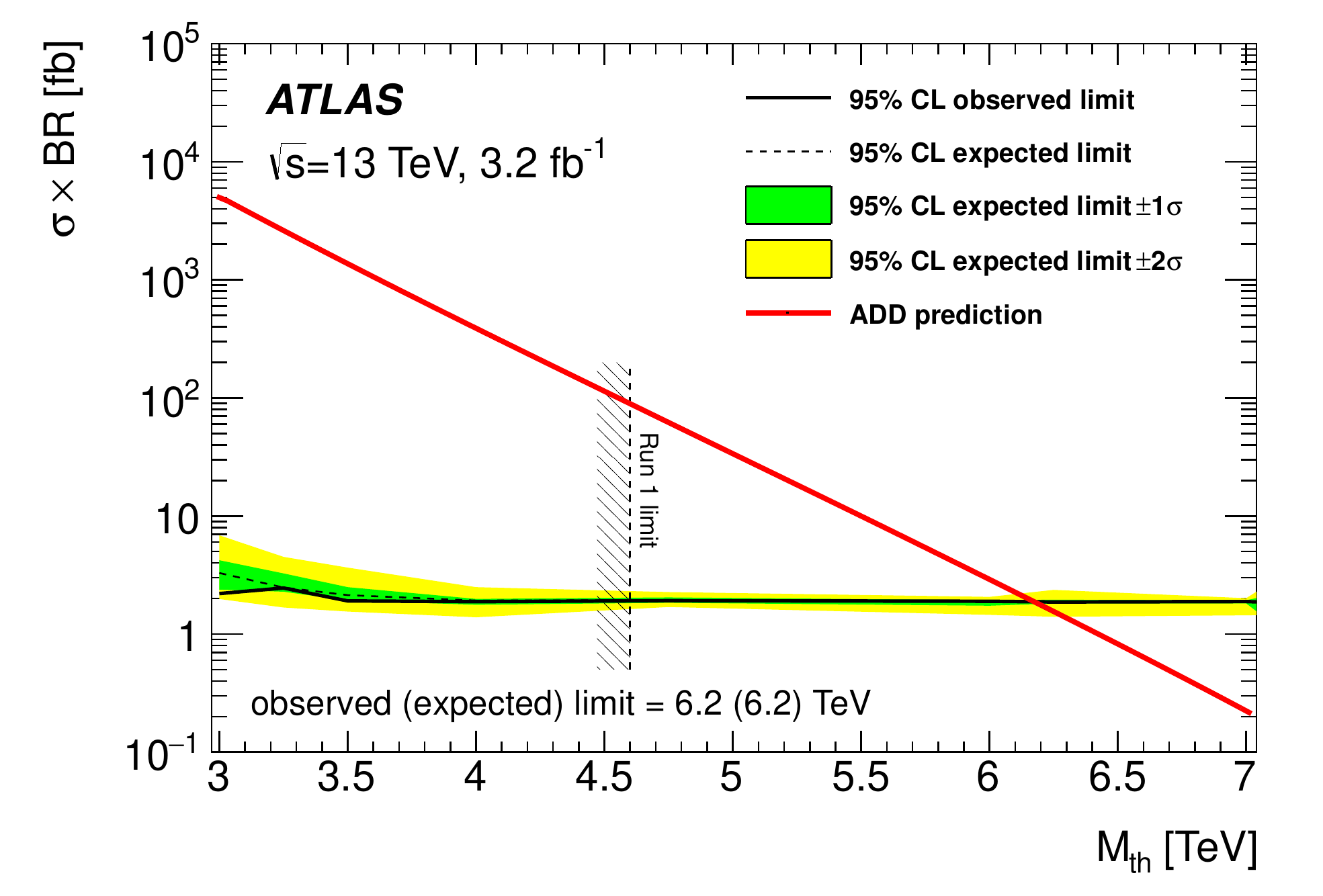}}
     \caption{
       Observed 95\% CL limits (dots and solid black line) on the
       production cross section times branching ratio to a photon
       and a quark or a gluon for
       (a) an excited quark $q^{*}$,
       (b) an RS1 ($n = 1$) QBH, and
       (c) an ADD ($n = 6$) QBH.
       The limits are shown as a function of the $q^*$ mass or
       the QBH production threshold mass.
       The median expected 95\% CL exclusion limits (dashed line),
       in the case of no expected signal, are also shown.
       The green and yellow bands correspond to the $\pm 1\sigma$
       and $\pm2\sigma$ intervals.
       The red solid lines show the predicted $\sigma \times BR$
       (at leading order in $\alpha_{\rm s}$ in
       the case of the $q^*$ model).
       The dashed red lines in (a) show how the PDF uncertainties
       affect the prediction.
       In the case of the $q^*$ and ADD QBH searches,
       the corresponding limits from the ATLAS Run 1 pp data are
       indicated (vertical line).}
     \label{fig:limit_data}
  \end{center}
\end{figure}

%-------------------------------------------------------------------------------

%-------------------------------------------------------------------------------
%\input{}
%-------------------------------------------------------------------------------

% All figures and tables should appear before the summary and conclusion.
% The package placeins provides the macro \FloatBarrier to achieve this.
% \FloatBarrier

%-------------------------------------------------------------------------------
\clearpage
\section{Conclusions}
\label{sec:conclusiona}

A search for phenomena beyond the Standard Model has been performed
using photon+jet events in \datalumi\ of proton--proton collisions
with a centre-of-mass energy of $\sqrt{s} = 13$~TeV
collected by the ATLAS detector at the Large Hadron Collider.
No significant excess in the \gj invariant mass distribution
was found with respect to a data-driven estimate of the
smoothly falling distribution predicted by the Standard
Model.

Limits at 95\%~CL using a profile likelihood method
are obtained for three signal processes.
They include
(i) generic signals yielding a Gaussian lineshape,
with intrinsic width between 2\% and 15\%
of the resonance mass;
(ii) excited quarks with vector-like couplings to
Standard Model particles and a compositeness scale equal to
their mass, and
(iii) non-thermal quantum black holes in a type-1 Randall--Sundrum model
with one extra spatial dimension, and in a Arkani-Hamed--Dimopoulous--Dvali
model with six extra spatial dimensions.
The limits on Gaussian-shaped resonances exclude
1.5~TeV resonances with visible cross sections above about 6 (50)~fb
and 5~TeV resonances with visible cross sections above about 0.8 (1.0)~fb
at $\sqrt{s}=13$~TeV for $\sigma_{\mathrm{G}}/M_{\mathrm{G}}=2\%$ (15\%).
Excited quarks are excluded for masses up to 4.4~TeV.
Non-thermal RS1 and ADD quantum black hole models are excluded for
masses up to 3.8~TeV and 6.2~TeV, respectively.

The limits on the excited quarks and the non-thermal quantum
black holes are the most stringent limits set to date in the
\gj final state and supersede the previous ATLAS results
at $\sqrt{s}$ = 8 TeV.

%-------------------------------------------------------------------------------

%-------------------------------------------------------------------------------
\section*{Acknowledgements}
%-------------------------------------------------------------------------------

% Acknowledgements for papers with collision data
% Version 15-Sep-2015

% Standard acknowledgements start here
%----------------------------------------------
We thank CERN for the very successful operation of the LHC, as well as the
support staff from our institutions without whom ATLAS could not be
operated efficiently.

We acknowledge the support of ANPCyT, Argentina; YerPhI, Armenia; ARC, Australia; BMWFW and FWF, Austria; ANAS, Azerbaijan; SSTC, Belarus; CNPq and FAPESP, Brazil; NSERC, NRC and CFI, Canada; CERN; CONICYT, Chile; CAS, MOST and NSFC, China; COLCIENCIAS, Colombia; MSMT CR, MPO CR and VSC CR, Czech Republic; DNRF, DNSRC and Lundbeck Foundation, Denmark; IN2P3-CNRS, CEA-DSM/IRFU, France; GNSF, Georgia; BMBF, HGF, and MPG, Germany; GSRT, Greece; RGC, Hong Kong SAR, China; ISF, I-CORE and Benoziyo Center, Israel; INFN, Italy; MEXT and JSPS, Japan; CNRST, Morocco; FOM and NWO, Netherlands; RCN, Norway; MNiSW and NCN, Poland; FCT, Portugal; MNE/IFA, Romania; MES of Russia and NRC KI, Russian Federation; JINR; MESTD, Serbia; MSSR, Slovakia; ARRS and MIZ\v{S}, Slovenia; DST/NRF, South Africa; MINECO, Spain; SRC and Wallenberg Foundation, Sweden; SERI, SNSF and Cantons of Bern and Geneva, Switzerland; MOST, Taiwan; TAEK, Turkey; STFC, United Kingdom; DOE and NSF, United States of America. In addition, individual groups and members have received support from BCKDF, the Canada Council, CANARIE, CRC, Compute Canada, FQRNT, and the Ontario Innovation Trust, Canada; EPLANET, ERC, FP7, Horizon 2020 and Marie Skłodowska-Curie Actions, European Union; Investissements d'Avenir Labex and Idex, ANR, Region Auvergne and Fondation Partager le Savoir, France; DFG and AvH Foundation, Germany; Herakleitos, Thales and Aristeia programmes co-financed by EU-ESF and the Greek NSRF; BSF, GIF and Minerva, Israel; BRF, Norway; the Royal Society and Leverhulme Trust, United Kingdom.

The crucial computing support from all WLCG partners is acknowledged
gratefully, in particular from CERN and the ATLAS Tier-1 facilities at
TRIUMF (Canada), NDGF (Denmark, Norway, Sweden), CC-IN2P3 (France),
KIT/GridKA (Germany), INFN-CNAF (Italy), NL-T1 (Netherlands), PIC (Spain),
ASGC (Taiwan), RAL (UK) and BNL (USA) and in the Tier-2 facilities
worldwide.
%----------------------------------------------

%-------------------------------------------------------------------------------
% \clearpage
% \appendix
% \part*{Appendix}
% \addcontentsline{toc}{part}{Appendix}
%-------------------------------------------------------------------------------

%-------------------------------------------------------------------------------
% If you use biblatex and either biber or bibtex to process the bibliography
% just say \printbibliography here
\printbibliography
% If you want to use the traditional BibTeX you need to use the syntax below.
%\bibliographystyle{bibtex/bst/atlasBibStyleWoTitle}
%\bibliography{PhotonIDPrerecommendations2015,bibtex/bib/ATLAS}
%-------------------------------------------------------------------------------

%-------------------------------------------------------------------------------
% Print the list of contributors to the analysis
% The argument gives the fraction of the text width used for the names
%-------------------------------------------------------------------------------
% \clearpage
% \PrintAtlasContribute{0.30}

%-------------------------------------------------------------------------------
% \clearpage
% \input{sections/auxiliary.tex}
%-------------------------------------------------------------------------------

%-------------------------------------------------------------------------------
 \clearpage
% ATLAS Collaboration author list
% Data extracted on 05-Dec-2015 for paper reference EXOT-2015-12
\begin{flushleft}
{\Large The ATLAS Collaboration}

\bigskip

G.~Aad$^\textrm{\scriptsize 85}$,
B.~Abbott$^\textrm{\scriptsize 112}$,
J.~Abdallah$^\textrm{\scriptsize 150}$,
O.~Abdinov$^\textrm{\scriptsize 11}$,
B.~Abeloos$^\textrm{\scriptsize 116}$,
R.~Aben$^\textrm{\scriptsize 106}$,
M.~Abolins$^\textrm{\scriptsize 90}$,
O.S.~AbouZeid$^\textrm{\scriptsize 136}$,
H.~Abramowicz$^\textrm{\scriptsize 152}$,
H.~Abreu$^\textrm{\scriptsize 151}$
% ...
\bigskip
\\
% ...
$^{11}$ Institute of Physics, Azerbaijan Academy of Sciences, Baku, Azerbaijan\\
% ...
$^{85}$ CPPM, Aix-Marseille Universit{\'e} and CNRS/IN2P3, Marseille, France\\
% ...
$^{90}$ Department of Physics and Astronomy, Michigan State University, East Lansing MI, United States of America\\
% ...
$^{106}$ Nikhef National Institute for Subatomic Physics and University of Amsterdam, Amsterdam, Netherlands\\
% ...
$^{112}$ Homer L. Dodge Department of Physics and Astronomy, University of Oklahoma, Norman OK, United States of America\\
% ...
$^{116}$ LAL, Univ. Paris-Sud, CNRS/IN2P3, Universit{\'e} Paris-Saclay, Orsay, France\\
% ...
$^{136}$ Santa Cruz Institute for Particle Physics, University of California Santa Cruz, Santa Cruz CA, United States of America\\
% ...
$^{150}$ Institute of Physics, Academia Sinica, Taipei, Taiwan\\
$^{151}$ Department of Physics, Technion: Israel Institute of Technology, Haifa, Israel\\
$^{152}$ Raymond and Beverly Sackler School of Physics and Astronomy, Tel Aviv University, Tel Aviv, Israel\\
% ...
\end{flushleft}

%-------------------------------------------------------------------------------

\end{document}